\documentclass[twocolumn,superscriptaddress,longbibliography,aps,pra,preprintnumbers]{revtex4-2}

\usepackage{graphicx}
\usepackage{bm}
\usepackage{color}
\usepackage{epstopdf}
\usepackage{amsmath}
\usepackage{amssymb}
\usepackage{epstopdf}
\usepackage{gensymb}
\usepackage{float}
\usepackage{ulem}

\begin{document}

\title{{Thermoelectric power factors of defective scandium nitride nanostructures\\ from first principles}}



\author{Luigi Cigarini}
\affiliation{IT4Innovations, V\v{S}B – Technical University of Ostrava, 17. listopadu 2172/15, 708 00 Ostrava, Czech Republic}

\author{Urszula Danuta Wdowik}
\affiliation{IT4Innovations, V\v{S}B – Technical University of Ostrava, 17. listopadu 2172/15, 708 00 Ostrava, Czech Republic}

\author{Dominik Legut}
\affiliation{Department of Condensed Matter Physics, Faculty of Mathematics and Physics, Charles University, Ke
Karlovu 3, 121 16 Prague 2, Czech Republic}
\affiliation{IT4Innovations, V\v{S}B – Technical University of Ostrava, 17. listopadu 2172/15, 708 00 Ostrava, Czech Republic}

\date{\today}

\begin{abstract}

{The thermoelectric properties of scandium nitride are strongly influenced by structural and electronic factors arising from defects and impurities. Nevertheless, the mechanisms by which these microscopic features affect transport are not yet fully understood. Experiments show a large variability in the electronic transport properties, with a strong dependence on the experimental conditions, and attempts to improve thermoelectric efficiency often lead to conflicting effects. In this work, we employ the Landauer approach to analyze the effects of different kinds of structural defects and impurities on electronic transport in scandium nitride. This approach allows us to relate the transport mechanisms to the structural and electronic modifications introduced in the lattice, with atomistic resolution. In light of these new insights, we propose a rationale relating part of the experimental variability to its microscopic origin.}



\end{abstract}

\maketitle

\section{Introduction} 

Scandium nitride (ScN) belongs to the group of transition metal nitrides (TMN), which have gained considerable interest as promising materials for thermoelectric applications \cite{C5TC03891J,NINGTHOUJAM201550,PhysRevMaterials.3.020301}. Today, the need for sustainable energy is well understood and multiple pathways are being explored to achieve this result. Techniques that minimize harmful by-products and energy waste in energy conversion processes deserve increasing consideration. Thermoelectric effects have significant potential to address these challenges, mainly by converting waste heat into reusable energy for industrial and domestic appliances \cite{Goldsmid,HE2024121813}. For these purposes, it is necessary to identify materials with excellent thermoelectric conversion properties, low cost, ease of processing, chemical stability, and no risks to human health. 

The efficiency in thermoelectric conversion of a material is expressed by its thermoelectric figure of merit, denoted as $zT$, and defined as: $zT = {\sigma S^2 T}/{\kappa}$, where $\sigma$ is the electrical conductivity, $S$ is the Seebeck coefficient, $T$ is the absolute temperature, and $\kappa$ is the total thermal conductivity (including electronic and lattice contributions) \cite{Goldsmid,10.1073/pnas.93.15.7436}.

TMNs are chemically and physically stable and are characterized by hardness and high melting points. Furthermore, these materials, in their solid states, are safe for human contact and exhibit high electrical conductivity and high charge carrier densities. ScN is a semimetal with an indirect band gap, reported by theoretical and experimental studies, of \( \sim0.9 \)~eV~\cite{PhysRevB.63.125119,PhysRevB.104.075118} and a carrier concentration ranging from \(10^{18}\)~cm\(^{-3}\) to \(10^{22}\)~cm\(^{-3}\)~\cite{PhysRevB.34.3876,PhysRevB.70.045303,HARBEKE1972335,DISMUKES1972365,oshima2014hydride,PhysRevB.104.075118,Burmistrova_2015,PhysRevB.109.155307}. The electron mobility of ScN is strongly dependent on the growth method and experimental conditions, and it has been reported in a wide range, from less than 10~cm\(^2\)/(V~\(\cdot\)~s)~\cite{moustakas1996growth} to 284~cm\(^2\)/(V~\(\cdot\)~s)~\cite{oshima2014hydride}. Early TMNs—with the only exceptions, to the best of our knowledge, of ScN and chromium nitride (CrN)—exhibit Seebeck coefficients in their bulk form, up to high temperatures, of the order \(-1\) to \(-10\)\(\,\mu\)V/K~\cite{ADACHI2005242,samsonov1962physical}.

Although these electronic properties make TMNs potentially suitable for thermoelectric applications, their overall performance in thermoelectric conversion remains insufficient for practical purposes, which led to their omission until recent years. 

ScN has a rigid face-centered cubic crystal structure—later described in section \ref{sub:stm} (rocksalt, similar to sodium chloride)—with an experimental lattice parameter of $\sim$~4.50~Å~\cite{LENGAUER1988412}. Its strong covalent bonds reduce phonon-phonon damping, resulting in a high lattice thermal conductivity, ranging between 10 and 12~\(\mathrm{W/(m \cdot K)}\) at room temperature (RT)~\cite{PhysRevB.97.085301}, which negatively affects the overall \(zT\) value in its bulk form.

The perspectives on ScN and TMNs in thermoelectricity changed significantly, however, when a remarkably high Seebeck coefficient—in absolute terms—was reported in ScN thin films grown using reactive magnetron sputtering on aluminum oxide (\(\text{Al}_2\text{O}_3\))~\cite{10.1063/1.3665945}. Shortly after, a similar observation was made for ScN thin films grown on magnesium oxide (MgO)~\cite{10.1063/1.4801886}. These samples, while exhibiting an increased absolute value of the Seebeck coefficient, reaching \(-156 \, \mu\mathrm{V/K}\) at 840~K~\cite{10.1063/1.4801886}, preserve high electrical conductivity, resulting at high temperatures in remarkable thermoelectric power factors \(\sigma S^2\) up to \(3.5 \times 10^{-3} \, \mathrm{W/(m \cdot K^2)}\) at 840~K~\cite{10.1063/1.4801886}.

These observations highlight the potential to enhance the thermoelectric performances of TMNs through nanostructuring with appropriate techniques. The strategies for improving the conversion efficiency of thermoelectric materials are inherently complex, as the \(zT\) value depends on properties such as electrical conductivity (\(\sigma\)), Seebeck coefficient (\(S\)), and thermal conductivity (\(\kappa\)), which are often strongly interdependent, but have opposing effects on \(zT\)~\cite{JIA2021100519,10.1073/pnas.93.15.7436}. Consequently, these effects strongly depend on the preparation conditions and are reflected on the properties of the resulting material.

Appropriate preparation techniques for ScN thin films can also significantly reduce their thermal conductivities, as observed with lithium (Li) ions irradiation during plasma-assisted molecular beam epitaxy \cite{doi:10.1021/acsaem.2c00485}, or by implanting argon (Ar) or atoms {or Helium (He) ions into ScN thin films during or after} reactive magnetron sputtering deposition using an ion beam implantation technique~\cite{doi:10.1021/acsaem.2c01672,bouteiller2025improving}. These procedures lead to a strong decrease in the in-plane thermal conductivity of the resulting films, reaching respectively 7 and 3~\(\mathrm{W/(m \cdot K)}\) at RT. Recently, a similar reduction has been reported on ScN thin films prepared by reactive magnetron sputtering, with periodic interruptions of the cathodic potential during the deposition procedure~\cite{MORECHEVALIER2025100674}. However, in addition to their impact on thermal transport, these techniques also affect electronic transport, leading to opposing effects—regarding thermoelectric efficiency—between the Seebeck coefficient and electrical conductivity. Specifically, they induce a significant increase—in absolute terms—of the Seebeck coefficient while causing a strong reduction in electrical conductivity. The total effect on the \(zT\) value is slightly positive, with an increase in the thermoelectric figure of merit primarily driven by the substantial reduction in thermal conductivity, despite a decrease in the thermoelectric power factor{~\cite{doi:10.1021/acsaem.2c00485,doi:10.1021/acsaem.2c01672,MORECHEVALIER2025100674,bouteiller2025improving}}.

{Theoretical and experimental studies have correlated the transport properties of ScN with its structural defects. The presence of chemical impurities and lattice imperfections has a strong impact on the physical properties of ScN even under room-temperature sputtering deposition, and an even larger one under standard high-temperature deposition conditions~\cite{CHOWDHURY2022101375}. Impurities and lattice vacancies have consequently been identified as key factors for thermoelectricity~{\cite{PhysRevB.86.195140,leFebvrier_2019,PhysRevB.99.161117,doi:10.1021/acs.nanolett.4c02920,PhysRevApplied.9.034019,10.1063/5.0230961,PhysRevB.110.115139}}.
However, the mechanisms through which these microscopic features regulate the thermoelectric properties remain not fully investigated. Understanding these microscopic regulation mechanisms is crucial for developing further optimization strategies capable of selectively tuning individual parameters without negatively affecting others.}

{In this work, we aim to achieve a deeper understanding of the microscopic mechanisms regulating electronic transport—and, consequently, thermoelectricity—in ScN.} We aim to do so by modeling the electronic transport in ScN nanowires (NW) and associating it with structural lattice defects, whose formation in the ScN lattice can be hypothesized under the described experimental conditions {\cite{10.1063/1.4801886,doi:10.1021/acsaem.2c00485,doi:10.1021/acsaem.2c01672,MORECHEVALIER2025100674,bouteiller2025improving}}. We find general conclusions applicable to ScN lattices regardless of their {dimensionality}. Specifically, we study point defects such as the inclusion of oxygen (O) impurities and atomic vacancies in place of the lattice nitrogen (N) atoms. Furthermore, we examine the effects of association of multiple defects in contiguous lattice positions and the effects of {associacion with stacking faults}.

\section{Method}

\subsection{The Landauer approach to quantum transport}
\label{sub:Theory}

\begin{figure}
  \includegraphics[width=\columnwidth]{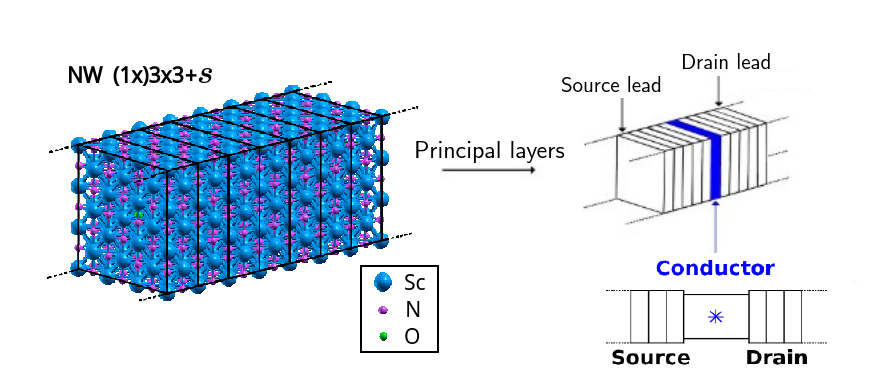}
  \caption{The Landauer model adapted to NW systems. The nanowire (NW) is divided into an infinite series of principal layers along its length, one layer acting as the central conductor and the other ones constituting the semi-infinite leads.}
  \label{fig:NW_Landauer}
\end{figure}

The Landauer model for quantum transport \cite{landauer1957spatial,landauer1987electrical,landauer1988spatial,datta1997electronic} relates either electric or thermal transport through a conducting system to the probability of carriers (electrons or phonons) being transmitted from a periodic semi-infinite source to a periodic semi-infinite drain (leads) across a central conductor acting as a scattering element. This model is particularly useful to link microscopic transport mechanisms to the measurable macroscopic properties.
Fisher and Lee \cite{PhysRevB.23.6851} introduced a nonequilibrium Green’s function (NEGF) formalism to relate the elements of the central conductor's scattering matrix to the Green’s functions of the leads and the central conductor, which describe how particles propagate through these systems under the influence of a force field. This approach yields a transmission function, \( \mathcal{T}(E) \), which describes the probability, as a function of the energy of the quantum carrier, that a particle is transmitted from the source to the drain while overcoming the scattering point \cite{datta1997electronic,PhysRevB.23.6851,PhysRevB.54.2600}:

\begin{equation}
\mathcal{T}(E) = \text{Tr} \left[ \Gamma_{s}(E) \mathcal{G}_{c}^r(E) \Gamma_{d}(E) \mathcal{G}_{c}^a(E) \right],
\end{equation}

where

\begin{equation}
\Gamma_{\{s,d\}}(E) = i \left( \Sigma_{\{s,d\}}(E) - \Sigma_{\{s,d\}}^\dagger(E) \right),
\end{equation}

\(\Sigma_{\{s,d\}}(E)\) and \(\Sigma_{\{s,d\}}^\dagger(E)\) are the retarded and advanced self-energies of the leads (source or drain), which contain the retarded and advanced Green's functions of the leads and the coupling matrices between the central conductor and the leads. Their rigorous expression can be found e.g., in Refs. \onlinecite{PhysRevB.60.7828,datta1997electronic,Camsari2023}. The retarded and advanced Green's functions of the central conductor, \(\mathcal{G}_{c}^r(E)\) and \(\mathcal{G}_{c}^a(E)\), describe the propagation of the electrons forwards and backwards in time in this section of the system. These can be straightforwardly calculated once knowing the real-space Hamiltonian of the system and the Green's functions of the semi-infinite leads \cite{PhysRevB.60.7828,Camsari2023,datta1997electronic}.

In this work, we calculate the Bloch space Hamiltonians of the systems using plane waves Density Functional Theory \cite{PhysRev.140.A1133} and the real-space Hamiltonians with a subsequent projection and filtering procedure \cite{PhysRevB.88.165127}. The Green's functions of the semi-infinite leads are then calculated dividing the real-space representation of the system into a series of principal layers orthogonal to the transport direction (see Fig. \ref{fig:NW_Landauer}) and then using the iterative Lopez-Sancho \textit{et al.} procedure \cite{LopezSancho_1985,PhysRevB.54.2600}.

Once obtained the energy dependent transmission function \( \mathcal{T}(E) \), this can be seen as the probability at given energy of electronic transmission through a single quantum channel of the conductor. The generalized Landauer formula defines the quantum conductance of single channels \( G^{(0)}(E) \) as \cite{datta1997electronic}:

\begin{equation}
G^{(0)}(E) = \frac{2e^2}{h} \mathcal{T}(E),
\end{equation}

where $e$ is the elementary charge and $h$ is the Planck constant.

The quantum conductance at Fermi level, \( G^{(0)}(E_F) \), corresponds to the zero-temperature conductance. The conductance at finite temperature is calculated, as a function of the chemical potential, as:

\begin{equation}
G(\mu) = \frac{2e^2}{h} L_0(\mu),
\end{equation}

where \( L_0 \) is the zeroth-order moment of transmission:

\begin{equation}
L_0 = \int_{-\infty}^{\infty} \mathcal{T}(E) \left( -\frac{\partial f_{T, \mu}(E)}{\partial E} \right) dE
\end{equation}

and $f_{T, \mu}(E)$ is the Fermi function at absolute temperature $T$ and chemical potential $\mu$.

The Seebeck coefficient (\( S \)) of the Landauer model is related to its electronic structure using the same moments of transmission formalism \cite{datta1997electronic,Goldsmid}:

\begin{equation}
S = -\frac{1}{eT} \frac{L_1}{L_0},
\end{equation}

where  \( L_1 \) is the first-order moment of transmission:

\begin{equation}
L_1 = \int_{-\infty}^{\infty} (E - \mu) \mathcal{T}(E) \left( -\frac{\partial f_{T, \mu}(E)}{\partial E} \right) dE.
\end{equation}

\subsection{Structural models}
\label{sub:stm}

As mentioned, ScN has a face-centered cubic (\textit{fcc}) rocksalt structure with a calculated lattice constant of 4.530~\AA. The primitive cell contains 2 atoms per cell, while it is possible to define a conventional simple cubic (\textit{sc}) cell using a larger crystal unit, containing 8 atoms per cell. The axes of the conventional \textit{sc} cell are aligned along the Sc–N bonds directions. 

{We chose to study the defective structures by embedding them in NW models. These nanostructures, being finite along the directions transverse to transport, allow us to obtain quantized, energy-dependent transmittance functions that can be directly compared with the electronic band structures of the systems, revealing important insights into the electronic transport mechanism that would not be accessible with 3D models.} Using periodic boundary conditions (PBC), we built the NW models along one of the Sc–N direction. The NWs were constructed by inserting sufficient space in the directions orthogonal to their length and replicating half of the surface layers of the obtained NW on the opposite sides (these additional layers are represented by \textit{+s} in the symbolism that we introduce to denote the NWs'geometries). This operation was done to make the cross-section of the NWs centrosymmetric. Two different periodicities were employed, corresponding to either one or three conventional cells along the NW direction, and different cross-section areas: (1x)1x1\textit{+s}, (1x)3x3\textit{+s}, (1x)5x5\textit{+s}, (3x)3x3\textit{+s}. These symbols describe the models' structures: in parentheses the length of the periodic unit—expressed in conventional cells—then the cross-sectional area, and the symbol \textit{+s} for additional layers.

Since we are interested in the effects of impurities and defects in ScN lattices regardless of their dimensionality, we first performed structural relaxation in three-dimensional bulk ScN systems and then constructed the defected NW models using the relaxed atomic positions obtained from three-dimensional bulk structures. Unless otherwise specified, all the results reported in this work regarding defected systems refer to relaxed structures.

We studied bulk-type configurated (BT) Landauer models, where the source and drain leads consist of a semi-infinite periodic repetition of principal layers identical to the central conductor. This configuration, with an infinite repetition of the defect, is useful for highlighting its effects on transport. It should be noted that in BT models with a periodicity of one conventional cell, structural defects located at the center of the periodic cell are structurally contiguous, being separated by only a single Sc atom. These systems allow for the study of defect association effects. In contrast, in systems with a periodicity of three conventional cells, such defects are more widely separated.

Next, we investigated two-leads (TL) Landauer models, in which a single central scattering region is considered and the source and drain leads are composed of principal layers of pure ScN. This second type of configuration is useful for providing a quantitative estimate of the presence of defects.

\begin{figure}
  \includegraphics[width=\columnwidth]{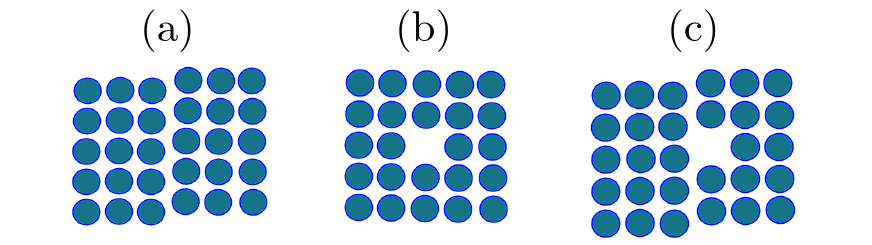}
  \caption{Schematic representations of (\textit{a}) a {stacking {fault}}, (\textit{b}) a point defect, and (\textit{c}) a combined configuration of a {planar {stacking fault}} and a point defect.}
  \label{fig:Symm_defects}
\end{figure}

The possible structural defects that may arise in the ScN lattice can be classified according to their symmetry into point defects and {planar {faults}}. Combinations of these two types of defects can also occur in spatially adjacent positions (Figure \ref{fig:Symm_defects}).

{We consider {stacking faults (\textit{sf})} perpendicular to the transport direction, in order to maximize the {effects of the structural mismatches} and observe them more clearly. We modeled these {planar defects} in the (1x)5x5\textit{+s} structures, associating them with point defects. After obtaining the relaxed NW structures with point defects, we replaced a relaxed atomic layer adjacent to the defect and perpendicular to the NW length, with a layer from the unrelaxed structure, centered along the same NW axis}.

Several types of point-type defects can be accounted: vacancies and substitutions of either Sc or N atoms, or clusters of multiple such defects. These defects can be ranked based on their relevance by calculating their formation enthalpy, revealing that substitutional N defects are the most energetically favorable and thus the most common in experimental samples \cite{BALASUBRAMANIAN201877}.

Concerning the chemical nature of defects, under magnetron sputtering deposition or molecular-beam epitaxy experimental conditions \cite{10.1063/1.3665945,10.1063/1.4801886,doi:10.1021/acsaem.2c00485,doi:10.1021/acsaem.2c01672,MORECHEVALIER2025100674}, the presence of oxygen impurities is likely unavoidable, even in ultra-high-vacuum (UHV) systems \cite{MORAM20088569,MORECHEVALIER2023156203,CICHON2024160867}. We studied substitutional defects where N sites are replaced either by structural vacancies or by oxygen atoms, and association of these {structures} with contiguous point defects or {planar mismatches}. We coherently compared each case to models of pure ScN nanowires of the same size.

\subsection{Detailed description}

\label{sub:detail}

The electronic structures of the systems were resolved by means of plane-wave (PW) Density Functional Theory (DFT) with periodic boundary conditions (PBC), as implemented in \textit{Quantum Espresso} (QE)~\cite{QE1,QE2,PhysRev.140.A1133}. We employed the Perdew-Burke-Ernzerhof (PBE) exchange-correlation functional, within the Generalized Gradient Approximation (GGA)~\cite{PhysRevLett.77.3865}. A Gaussian smearing function with a degauss parameter of 0.018 Ry was applied to the electronic occupation function. The atomic nuclei and core-valence electrons were represented within the pseudopotential approximation using projected augmented wavefunctions (PAW)~\cite{PhysRevB.50.17953}.

The atomic positions in systems with structural point defects were obtained through structural relaxation calculations performed using the Broyden–Fletcher–Goldfarb–Shanno (BFGS) algorithm \cite{head1985broyden}. These structural relaxations were carried out on bulk systems, constructed with PBC. We introduced the structural defect at the center of supercells of size 3x3x3 and 1x5x5 conventional cells. For the structural relaxation calculations, the eigenvalues were calculated on uniform 3x3x3 k-points grids spanning the whole first Brillouin zone of the systems{, exploiting space-group symmetries where possible}; we employed a PW basis set defined by a cutoff energy of \(40 \, \text{Ry}\); the electronic charge density was recalculated after each diagonalization using a larger set of plane waves, defined by a cutoff energy of \(320 \, \text{Ry}\), before proceeding to the next step in the self-consistent field (SCF) cycle. The convergence thresholds were set to \(10^{-6}\) a.u. for the total energy and \(10^{-5}\) a.u. for the forces.

After obtaining the atomic positions, as modified by structural defects, the defected and pristine NW systems were constructed as explained in the previous subsection (\ref{sub:stm}). The electronic structures of the NW systems were obtained using a PW basis set defined by a cutoff energy of \(50 \, \text{Ry}\) and recalculating the electronic charge density after each diagonalization step in the SCF cycle using a larger cutoff energy of \(400 \, \text{Ry}\). The eigenvalues were calculated on a uniform k-points row spanning the whole first Brillouin zone of the systems only along the NW direction: 7 k-points for (1x)5x5\textit{+s} systems and 5 k-points for (3x)3x3\textit{+s} systems.

By means of the \textit{projwfc} code included in QE~\cite{QE1,QE2}, the resulting electronic structure, defined in a Bloch space, has been projected onto a real-space basis set composed of orthogonalized atomic wavefunctions. After determining a projectability parameter for each electronic eigenstate as implemented in the \textit{WanT} code \cite{WanT,PhysRevB.88.165127}, a subsequent filtering and shifting procedure has been applied using a filtering threshold of 0.8 on the projectability parameter. We applied a shift of 5.5 eV to the discarded eigenstates in order to remove them from the vicinity of the Fermi level and cancel their effect on the subsequent calculations. The remaining eigenstates constituted the real-space electronic structure that has been used for the NEGF calculations.

The electron transmission function, \(\mathcal{T}(E)\), was calculated using the \textit{WanT} code \cite{WanT}. The electronic conductivities and Seebeck coefficients were calculated as described in subsection \ref{sub:Theory}.

\section{Results and discussion}

{\subsection{Electronic band structures}}

\begin{figure*}[th]
\centering\includegraphics[width=\textwidth]{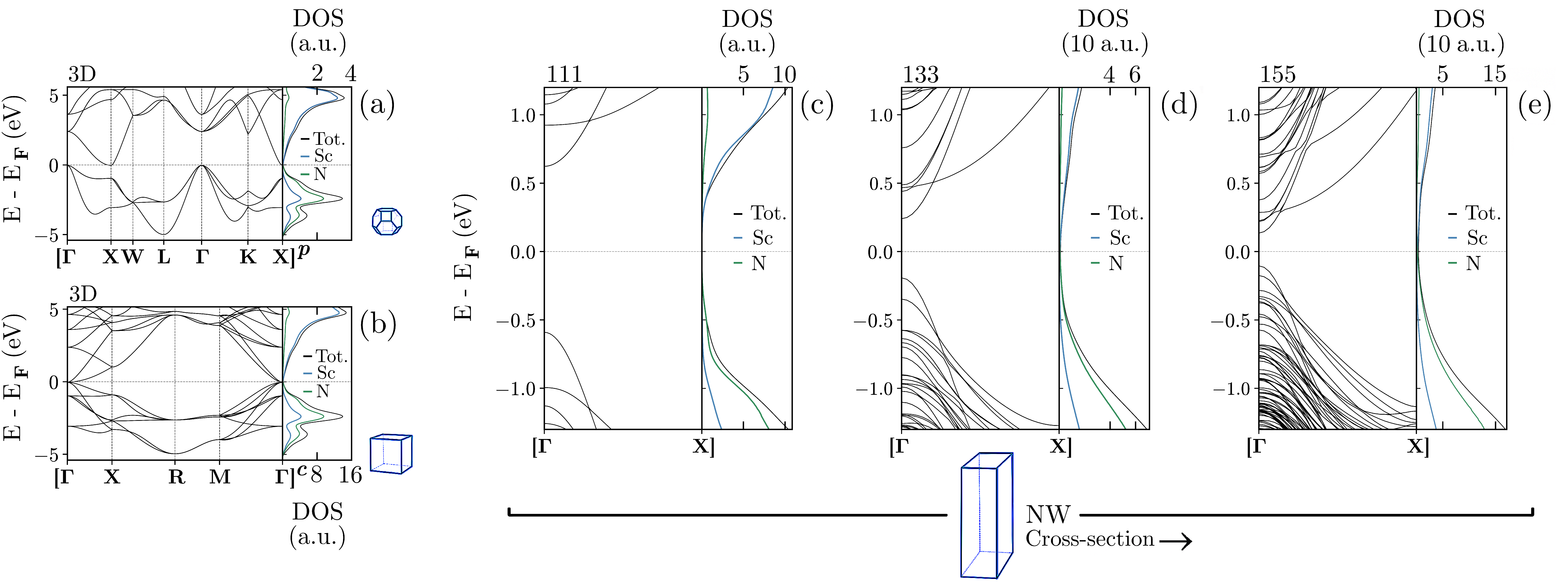}
 \caption{Electronic structures of pure ScN three-dimensional (3D) and NW systems. Panels \textit{a} and \textit{b}: Calculated electronic band structures and densities of states (DOS) of pure ScN in its 3D bulk form, represented along the high-symmetry lines of (panel \textit{a}) the primitive (\textit{p}) face-centered cubic system (2 atoms per cell) and (panel \textit{b}) a conventional (\textit{c}) simple cubic system (8 atoms per cell). Panels \textit{c}, \textit{d}, and \textit{e}: Calculated electronic band structures represented along the NW direction and DOS of pure ScN NW systems, as described in the text: (\textit{c}) (1×)1×1\textit{+s}, (\textit{d}) (1×)3×3\textit{+s}, and (\textit{e}) (1×)5×5\textit{+s}. To simplify the panels, we use the concise symbols (\textit{c}) 111, (\textit{d}) 133, and (\textit{e}) 155, respectively. The shapes of the Wigner-Seitz cells of the different referred lattices are schematically presented in blue.}
 \label{fig:bands_pure}
\end{figure*}

In Figure~\ref{fig:bands_pure}, we present the calculated electronic band structures of pure ScN represented along high-symmetry lines of the primitive (\textit{p}) face-centered cubic (\textit{fcc}) lattice—containing two atoms per unit cell—and the conventional (\textit{c}) simple cubic (\textit{sc}) lattice—containing eight atoms per cell {\footnote{The electronic band structures of the bulk three-dimensional systems were calculated as described in section \ref{sub:detail}, using an 81×81×81 k-point grid spanning the first Brillouin zone of the primitive cell and a 21×21×21 k-point grid for the simple cubic conventional cell, exploiting space-group symmetries where possible to reduce the computational effort.}}. The indirect band gap [\(\Gamma\text{--}X\)]\textsuperscript{\textit{p}} observed in the \textit{fcc} representation appears as a direct gap at the \(\Gamma\) point in the \textit{sc} representation, due to band folding at the \(\Gamma\) point. Theoretical literature reports an indirect [\(\Gamma\text{--}X\)]\textsuperscript{\textit{p}} band gap of \( \sim0.9 \)~eV for pure ScN, calculated using quasiparticle corrections \cite{PhysRevB.74.245208,PhysRevB.62.13538}, exact-exchange approach \cite{PhysRevB.63.125119} or hybrid functionals~\cite{PhysRevB.91.045104,PhysRevB.104.075118}. These theoretical values are in good agreement with experimental literature~\cite{DISMUKES1972365,Porte_1985,PhysRevB.63.125119}. This indirect band gap, however, is not captured by DFT, due to the well-known underestimation of the fundamental gap. In either the local density approximation (LDA) or GGA (Figure \ref{fig:bands_pure}), it is predicted to be zero~\cite{AZOUAOUI2021106090}. Apart from this known issue, the electronic structure computed using DFT does not exhibit significant topological or analytic variations in the shape of the bands near to the Fermi energy when compared to those obtained using quasiparticle corrections, exact exchange or hybrid functionals \cite{PhysRevB.74.245208,PhysRevB.86.195140}.

Within our theoretical approach, it is convenient to model NW systems. General conclusions, applicable also to larger systems, can be derived from these ScN models. One of the effects of nanodimensionality, compared to the bulk system, is the opening of the gap at the \(\Gamma\) point, as shown in Figure~\ref{fig:bands_pure}.

The studied NW systems are constructed in the ScN structure along one of the directions of the Sc–N bonds ([\(\Gamma\text{--}X\)]\textsuperscript{\textit{c}}). In Figure~\ref{fig:bands_pure}, we report the electronic band structures of pristine NW systems, along their axial \(\Gamma\text{--}X\) direction. The coordinate system that we use for NWs shares the same orientation—with respect to the ScN crystal—as that of the described \textit{sc} cell. In Figure~\ref{fig:bands_pure}, we compare three systems with cross-sections corresponding to 1 conventional cell: (1x)1x1\textit{+s}, 9 conventional cells: (1x)3x3\textit{+s}, and 25 conventional cells: (1x)5x5\textit{+s}. The fundamental gap at the \(\Gamma\) point decreases from 1.21 eV in the (1x)1x1\textit{+s} system to 0.43 eV in (1x)3x3\textit{+s} and to 0.24 eV in (1x)5x5\textit{+s}.


\begin{figure}
  \includegraphics[width=\columnwidth]{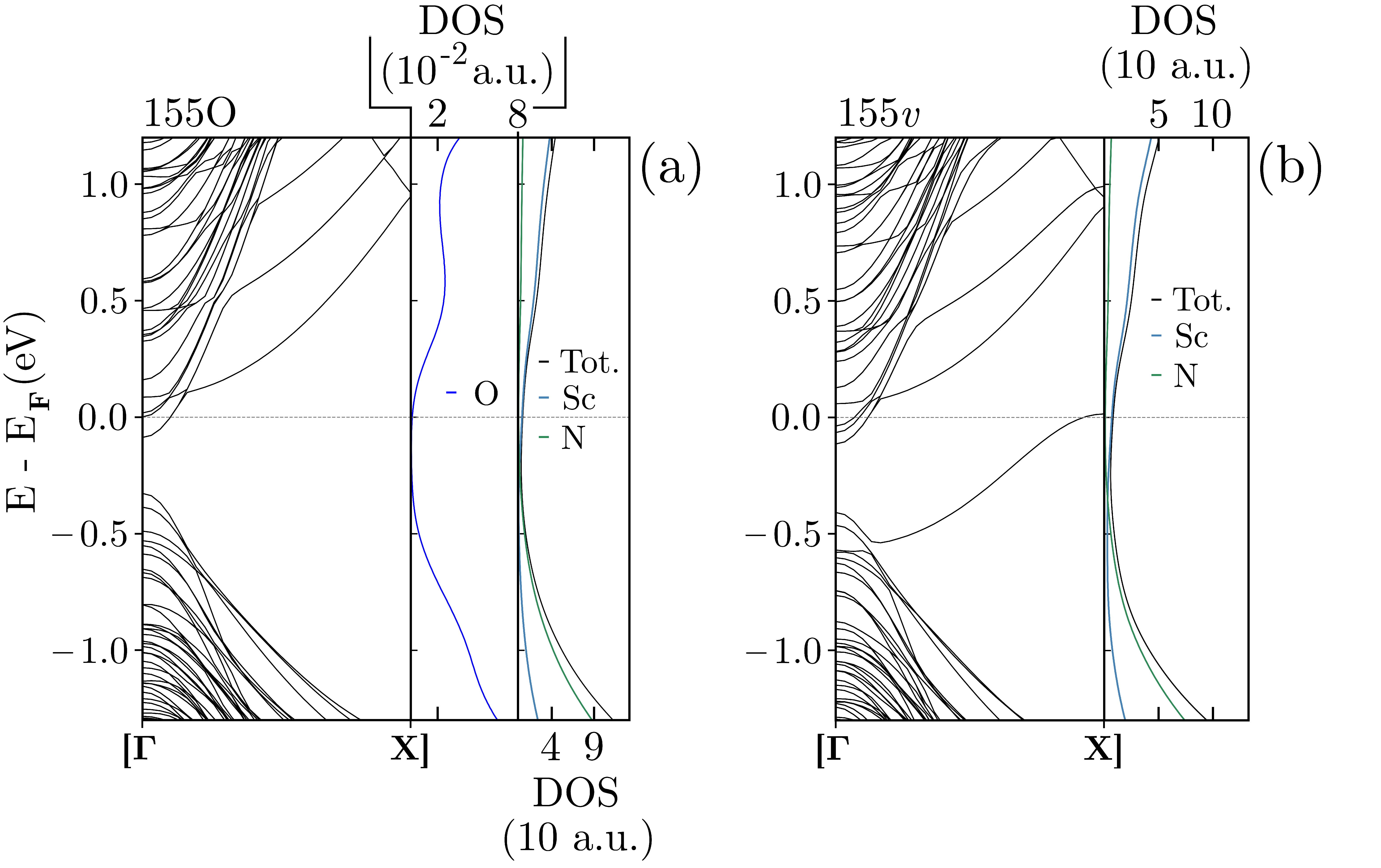}
  \caption{Electronic band structures along the axial direction of defected (1×)5×5\textit{+s} ScN NW systems. Panel \textit{a}: with an O atom replacing the central N site (155O). Panel \textit{b}: with a vacancy replacing the central N site (155\textit{v}).}
  \label{fig:bands_155O_155v}
\end{figure}

\begin{figure}
  \includegraphics[width=\columnwidth]{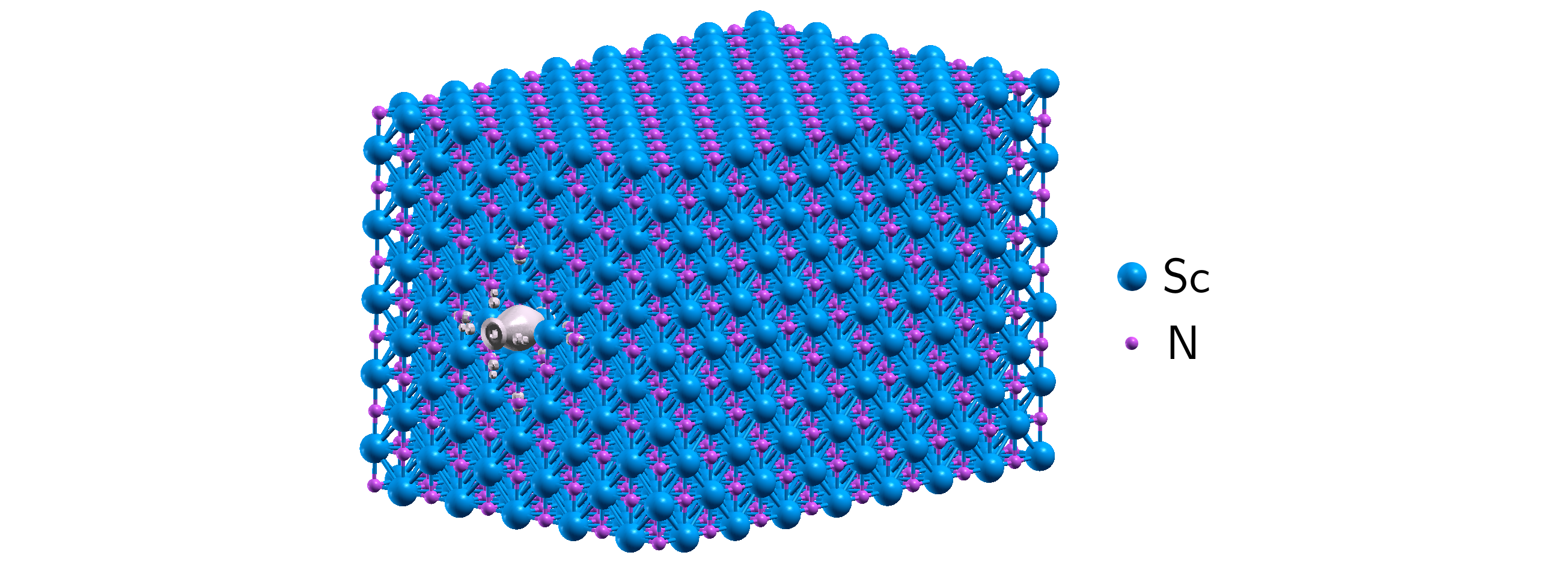}
  \caption{A section of the 155\textit{v} NW system. In gray: real-space representation at the $\Gamma$-point of the wavefunction associated with the electronic state whose band connects the points ($\Gamma$, $-0.58$) and (X, $0.01$) in the 155\textit{v} electronic structure (Fig.~\ref{fig:bands_155O_155v}, panel \textit{b}). The wavefunction is represented as its contribution to the electronic charge density at an isovalue of 0.003 a.u.{, using the \textit{XCrySDen} visualization tool \cite{Kokalj1999}.}}
  \label{fig:155v_band}
\end{figure}

\begin{figure}
  \includegraphics[width=\columnwidth]{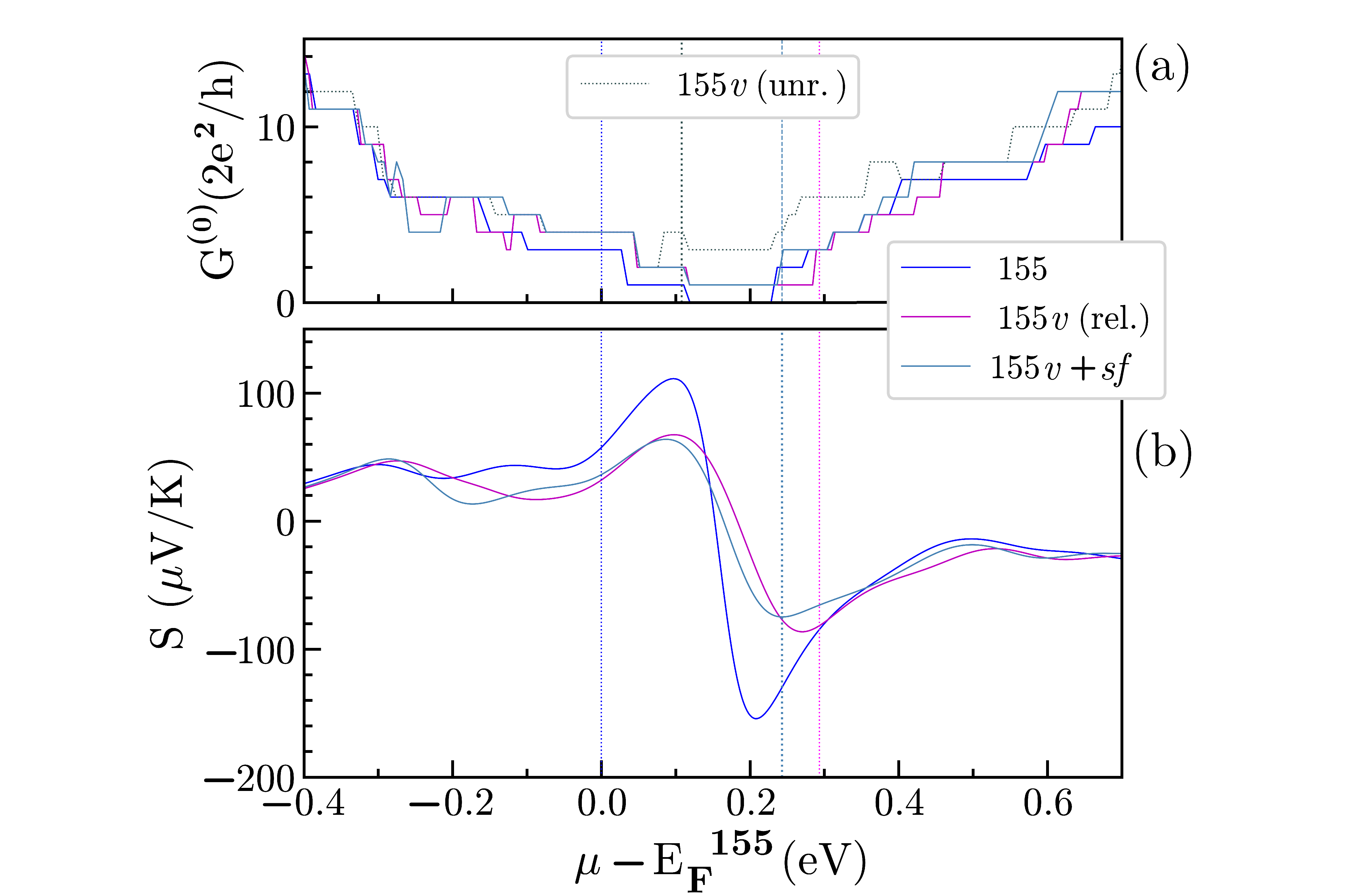}
  \caption{Panel \textit{a}: Calculated electronic conductances at 0~K as functions of the chemical potential of (1x)5x5\textit{+s} ScN NW systems: pure (155); BT defected system with a vacancy replacing the central N atom (155\textit{v}): relaxed and unrelaxed; and a BT defected system in which a {stacking fault} is contiguous to the N site vacancy and perpendicular to NW (155\textit{v}+{\textit{{sf}}}). Panel \textit{b}: Calculated Seebeck coefficients as functions of the chemical potential at 400 K for the three systems 155, BT 155\textit{v} (relaxed) and BT 155\textit{v}+{\textit{{sf}}}. Vertical dashed lines represent the calculated Fermi levels of the systems.}
  \label{fig:cond_Seebeck_155v}
\end{figure}

In Figure~\ref{fig:bands_155O_155v}, we compare the electronic band structures of the defected (1×)5×5\textit{+s} systems with central oxygen atoms and vacancies. While the presence of oxygen only leads to an $n$-type doping effect, with a rigid shift of the Fermi level without significantly distorting the band topology, the system with vacancies, in addition to a similar effect, exhibits also a strong distortion in the shape of a single band---whose real-space representation is shown in Figure \ref{fig:155v_band}---to the point that it crosses the Fermi level. This, in turn, affects the profile of the electronic transmission function (Figure \ref{fig:cond_Seebeck_155v}), with subsequent effect on the Seebeck coefficient. {In order to exclude the possibility that this feature results from nanostructuring effects in the NW model, we calculated the electronic band structure of the same systems by removing the additional \textit{+s} layers and reducing the cell dimensions along the directions transverse to the NW length, thereby restoring the 3D periodicity. As a result, we confirm that this feature is also present in this case, demonstrating that it is not a consequence of the reduced dimensionality.}

{\subsection{Seebeck coefficients}}

As seen in previous theoretical literature, the measurable electronic parameters of thermoelectricity—namely the Seebeck coefficient and electric conductivity—can be calculated from DFT, yielding satisfactory agreement with experimental values even without introducing any corrections beyond a rigid shift of the bands~\cite{10.1063/1.4801886,AZOUAOUI2021106090,MORECHEVALIER2025100674}.

\begin{figure}
\includegraphics[width=\columnwidth]{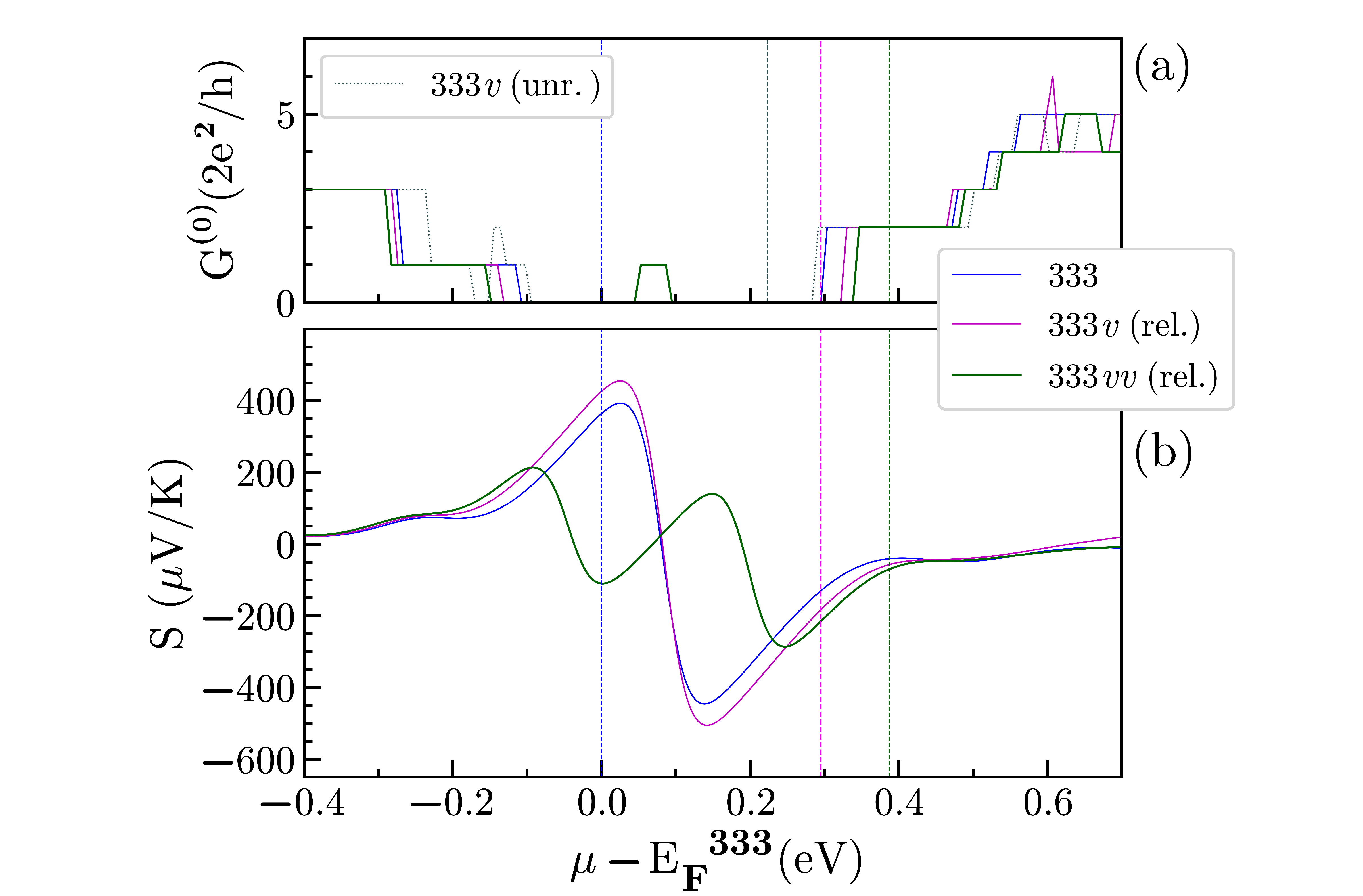}
  \caption{Panel \textit{a}: Calculated electronic conductances at 0~K as functions of the chemical potential of 3x(3x3)\textit{+s} ScN NW systems: pure (333); BT defected system with a vacancy replacing the central N atom (333\textit{v}): relaxed and unrelaxed; BT defected system with two vacancies replacing two N atomic sites along the central axis of the NW (relaxed: 333\textit{vv}). Panel \textit{b}: Calculated Seebeck coefficients as functions of the chemical potential at 400 K for the systems 333, BT 333\textit{v} relaxed and BT 333\textit{vv} relaxed. Vertical dashed lines represent the calculated Fermi levels of the systems.}
  \label{fig:cond_Seebeck_333v}
\end{figure}

\begin{figure}
  \includegraphics[width=\columnwidth]{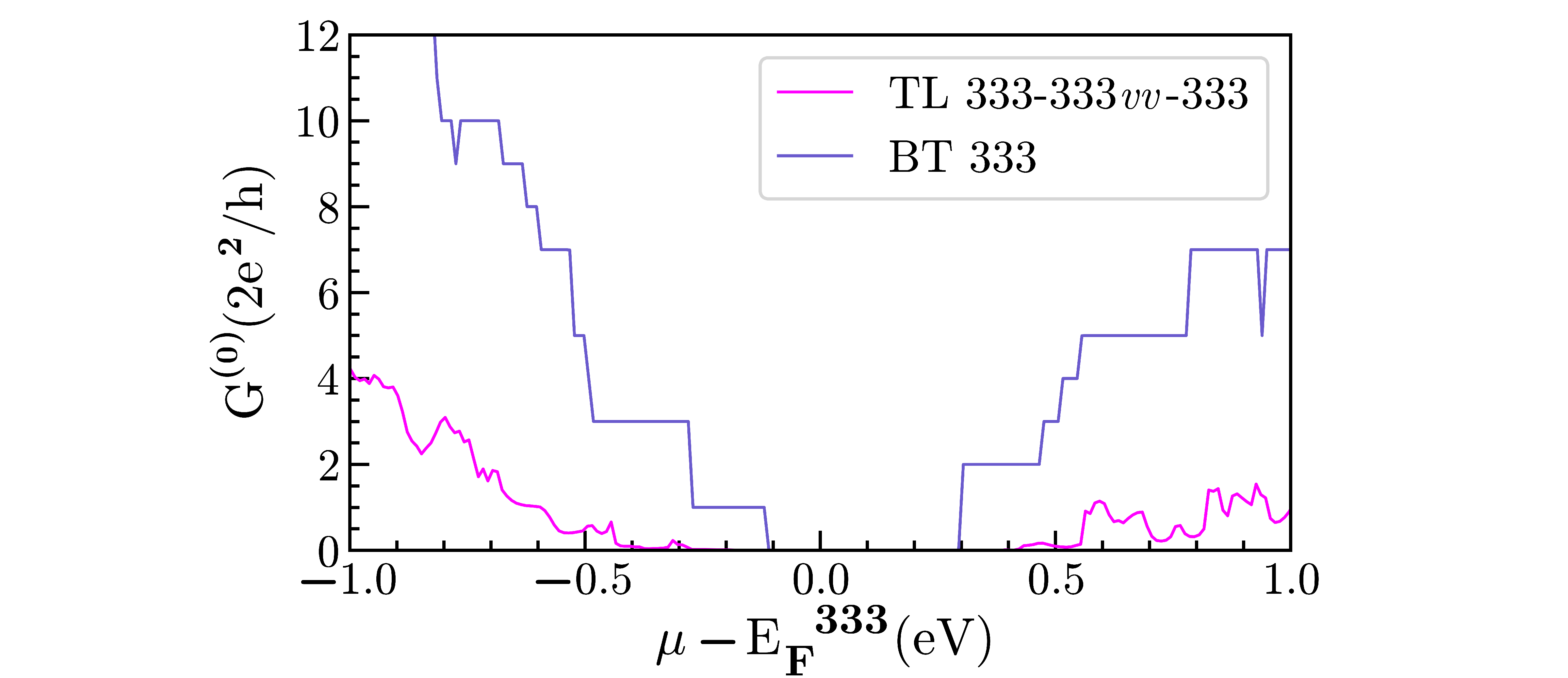}
  \caption{Comparison of the calculated conductances at 0~K as functions of the chemical potential for the 3×(3×3)\textit{+s} ScN NW models: pristine BT 333 and TL model composed by 333\textit{vv} as central conductor and pristine 333 as leads (333-333\textit{vv}-333).}
  \label{fig:TL_cond_333}
\end{figure}

\begin{figure}
  \includegraphics[width=\columnwidth]{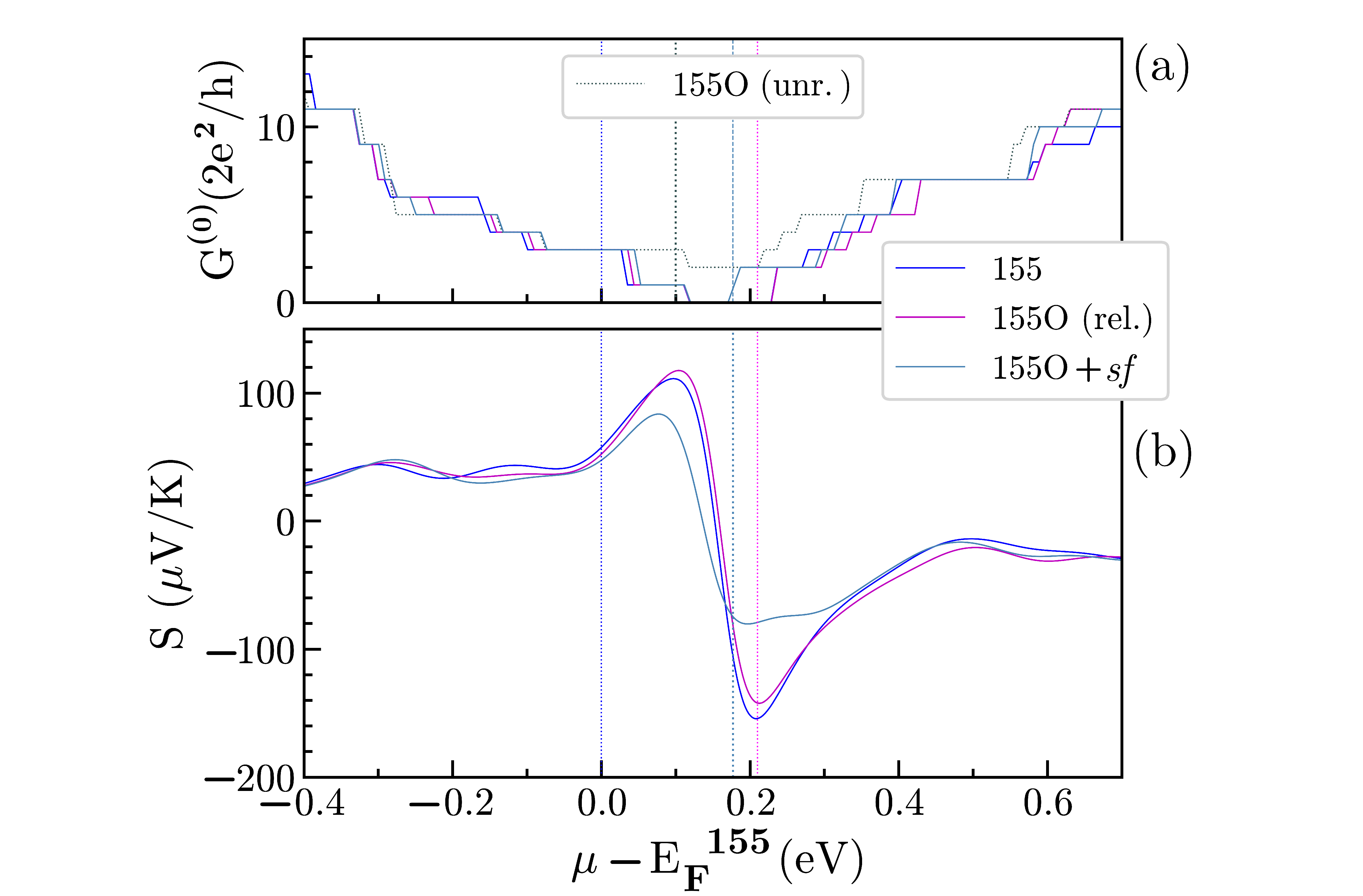}
  \caption{Panel \textit{a}: Calculated electronic conductances at 0~K as functions of the chemical potential of (1x)5x5\textit{+s} ScN NW systems: pure (155); BT defected system with an O atom replacing the central N site (155O): relaxed and unrelaxed; and a BT defected system in which a stacking {fault} is contiguous to the central O impurity and perpendicular to NW (155O+{\textit{{sf}}}). Panel \textit{b}: Calculated Seebeck coefficients as functions of the chemical potential at 400 K for the three systems 155, BT 155O (relaxed) and BT 155O+{\textit{{sf}}}. Vertical dashed lines represent the calculated Fermi levels of the systems.}
  \label{fig:cond_Seebeck_155O}
\end{figure}

\begin{figure}
  \includegraphics[width=\columnwidth]{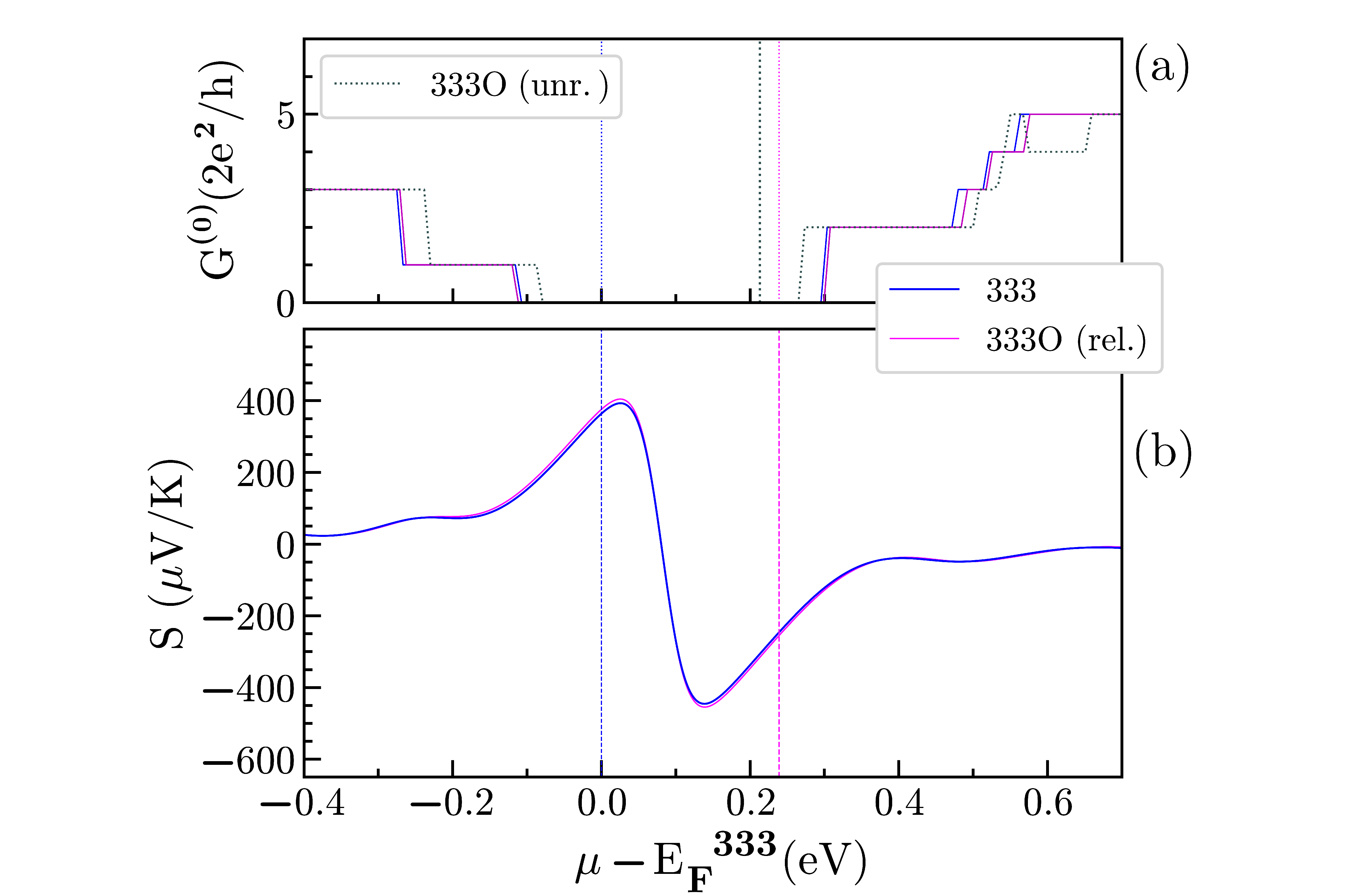}
  \caption{Panel \textit{a}: Calculated electronic conductances at 0~K as functions of the chemical potential of 3x(3x3)\textit{+s} ScN NW systems: pure (333); BT defected system with an O atom replacing the central N position (333O): relaxed and unrelaxed. Panel \textit{b}: Calculated Seebeck coefficients as functions of the chemical potential at 400 K for the systems 333 and BT 333O relaxed. Vertical dashed lines represent the calculated Fermi levels of the systems.}
  \label{fig:cond_Seebeck_333O}
\end{figure}

\begin{figure}
  \includegraphics[width=\columnwidth]{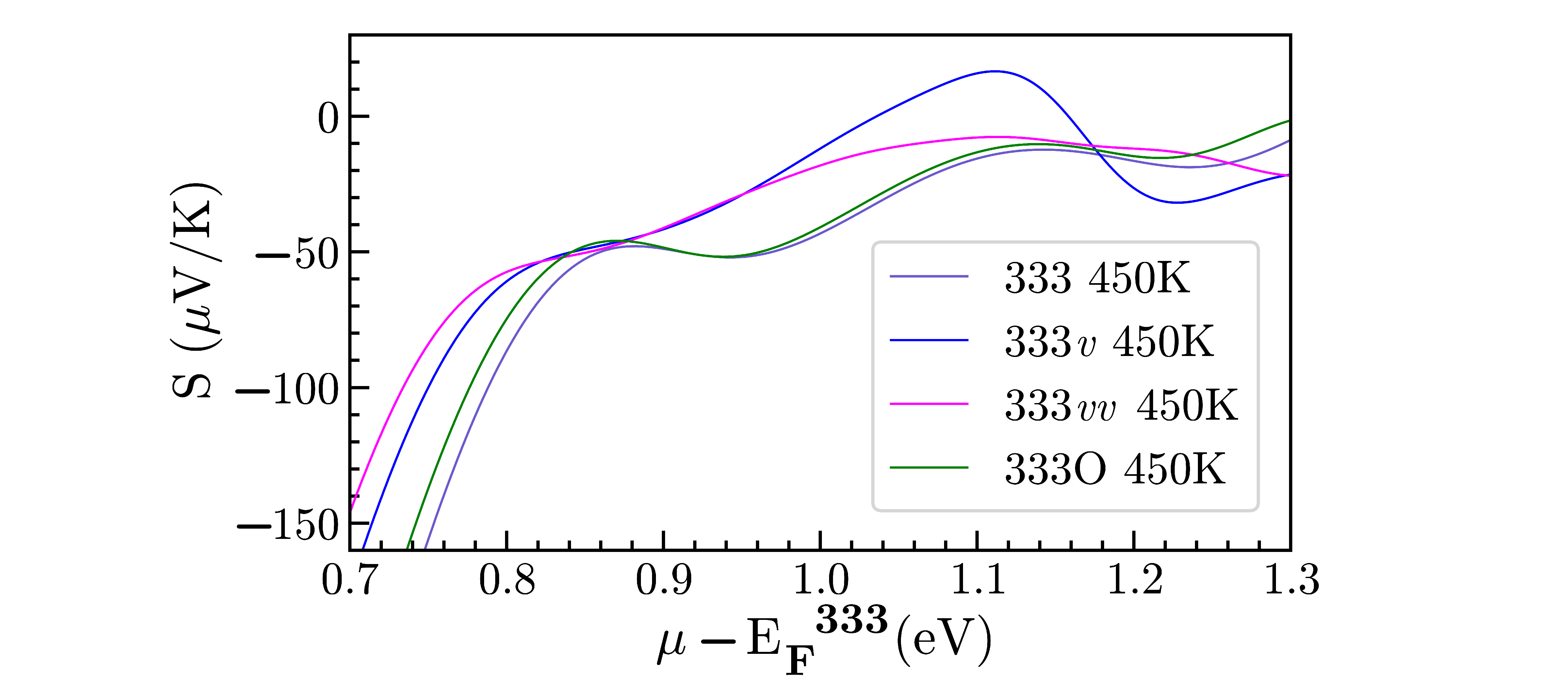}
  \caption{Seebeck coefficients calculated as functions of the chemical potential in BT 3×(3×3)\textit{+s} ScN NW systems at 450 K. The figure presents the results for the pristine system and three defected systems: with a single vacancy at an N site (333\textit{v}), with a single atomic O impurity (333O), and with a double vacancy at N sites (333\textit{vv}).}
  \label{fig:333_Seebeck}
\end{figure}

\begin{figure}
  \includegraphics[width=\columnwidth]{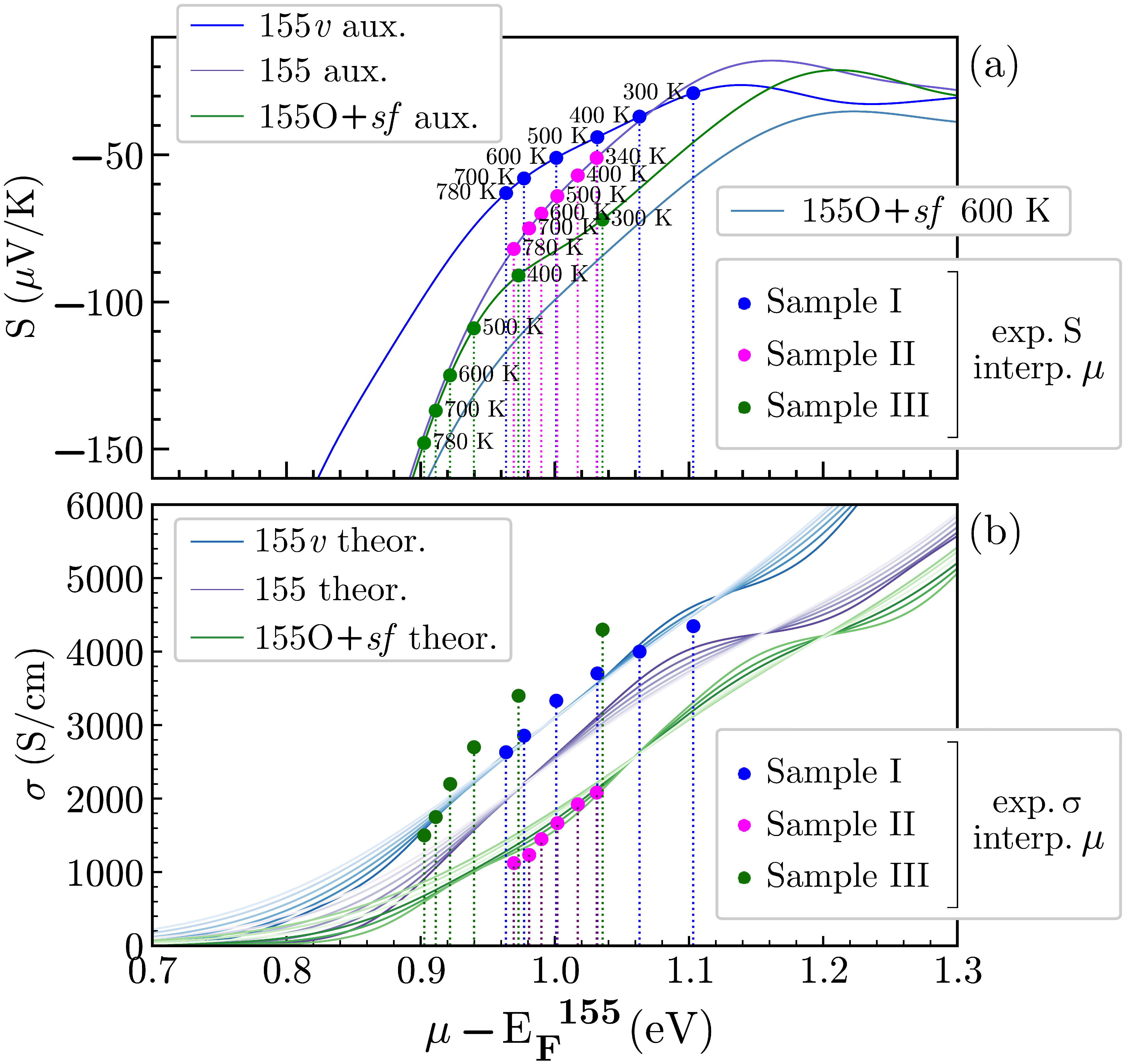}
  \caption{Panel \textit{a}: Auxiliary functions employed to relate the experimental Seebeck coefficients to theoretical values of chemical potential. The figure reports the functions of pristine and two different BT defected (1x)5x5\textit{+s} ScN NW structures (\textit{v}+\textit{v} and O+{\textit{{sf}}}). These functions are calculated as the Seebeck functions at 450 K. The figure also reports, for a comparison, the Seebeck function calculated at 600 K for the BT 155O+{\textit{{sf}}} structure. Panel \textit{b}: The lines represent calculated electric conductivities as functions of the chemical potential for the same three model systems, at different temperatures. Color intensities of the lines increase with temperature from 300 to 780 K. The points represent, at their calculated chemical potentials, experimental values of electric conductivities: Sample I: ScN from Ref. \onlinecite{MORECHEVALIER2025100674}, Sample II: ScN-T from Ref. \onlinecite{MORECHEVALIER2025100674} and Sample III: ScN from Ref. \onlinecite{10.1063/1.4801886}.}
  \label{fig:155_model}
\end{figure}

Within our theoretical approach, we obtain quantitative results by consistently comparing coherent systems, e.g. a system containing an impurity with the corresponding pristine system of the same dimensionality. When we compare energy dependent functions—such as transmission, Seebeck coefficient and conductance of corresponding systems (namely in Figures \ref{fig:cond_Seebeck_155v}, \ref{fig:cond_Seebeck_333v}-\ref{fig:155_model})—we shift them on the energy scale to ensure structural correspondence. Specifically, this is best achieved in (1×)5×5\textit{+s} systems when the transmittances of the quantum channels associated with the highest-energy valence bands are aligned, and in (3×)3×3\textit{+s} systems when the central values of the fundamental gaps are matched at the same energy.


In Figure~\ref{fig:cond_Seebeck_155v} (panel~\textit{a}), we compare the calculated transmittance functions for the (1×)5×5\textit{+s} systems in pure (155) and BT defected configurations, containing central vacancies (155\textit{v}). 
The same kind of defect is examined in BT (3×)3×3\textit{+s} models in  Figure \ref{fig:cond_Seebeck_333v}.
We observe a significant reduction in the profile of the Seebeck function in the systems with the shortest periodicity (155\textit{v}, Figure~\ref{fig:cond_Seebeck_155v}, panel~\textit{b}). The energy-dependent transmission function of the BT 155\textit{v} systems differs notably from that of the corresponding pure system (155), remaining finite in the region from 0.1 to 0.3~eV and exhibiting lower values in the range from 0.2 to 0.35~eV (Figure~\ref{fig:cond_Seebeck_155v}, panel~\textit{a}). These differences between transmission functions underlie the different profiles of the Seebeck coefficient functions and are largely due to structural modifications induced by the relaxation of the lattice in the presence of vacancies, as visible when comparing the transmission functions of relaxed and unrelaxed system in Figure~\ref{fig:cond_Seebeck_155v} (panel~\textit{a}). 

\begin{figure}
  \includegraphics[width=\columnwidth]{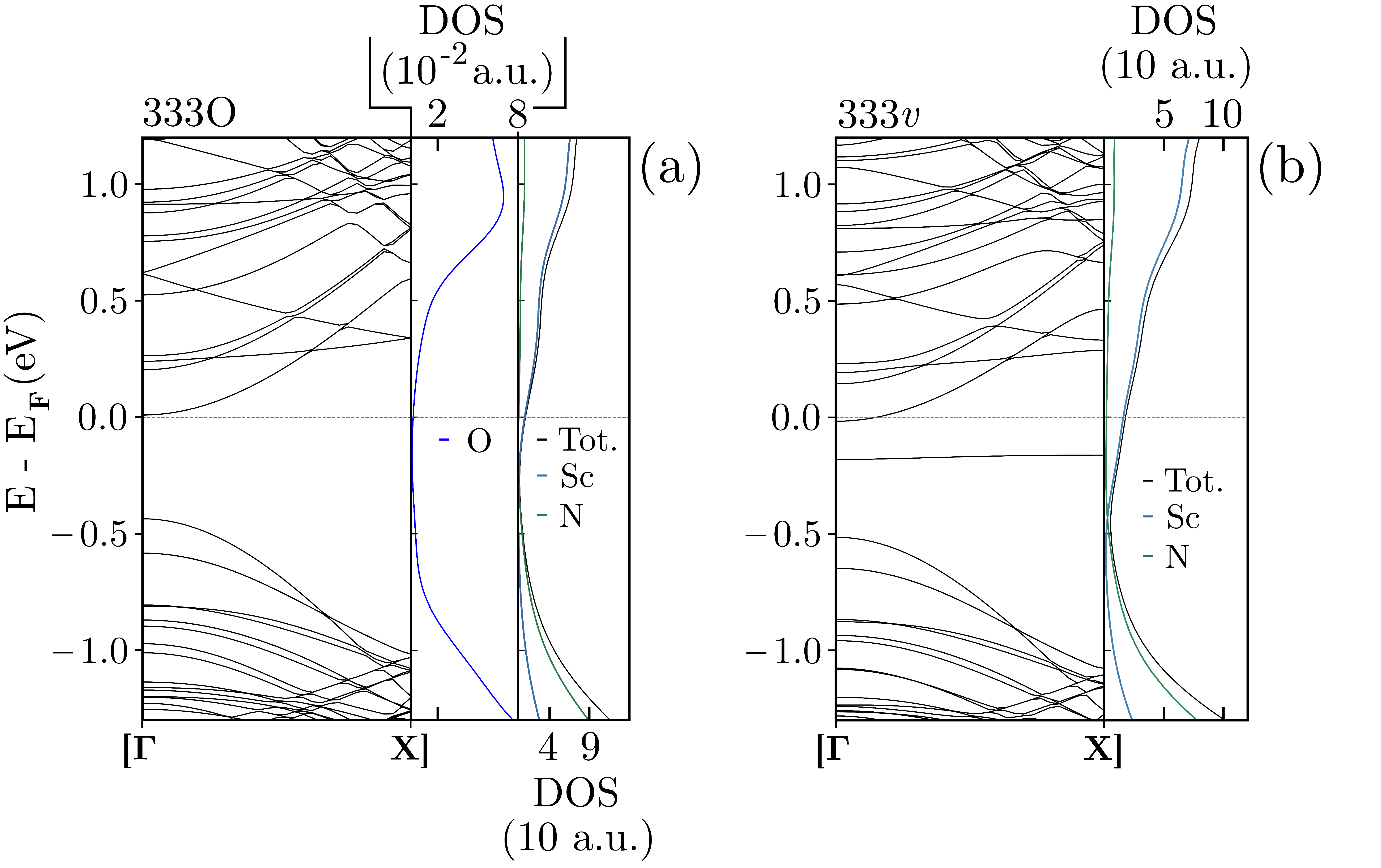}
  \caption{Electronic band structures along the axial direction of defected (3×)3×3\textit{+s} ScN NW systems. Panel \textit{a}: with an O atom replacing the central N site (333O). Panel \textit{b}: with a vacancy replacing the central N site (333\textit{v}).}
  \label{fig:bands_333O_333v}
\end{figure}

In the case of isolated central vacancies in BT systems (333\textit{v}, Figure \ref{fig:cond_Seebeck_333v}), no reduction in the Seebeck coefficient profile is observed. On the contrary, the presence of an isolated vacancy results in a slightly increased absolute value of the peak maximum. The calculated band structure of the 333\textit{v} system, shown in Figure \ref{fig:bands_333O_333v}, presents a flat, non-conductive band within the fundamental gap below the Fermi level, in place of the conductive band observed before in the 155\textit{v} system.

A reduction in the absolute value of the Seebeck coefficient is instead observed when two vacancies are introduced along the central axis in the direction of the nanowire in the (3×)3×3\textit{+s} cell (333\textit{vv}). In this configuration, the vacancies are contiguous and separated by only one Sc atom, similarly to what occurs in the BT 155\textit{v} system. This result confirms that a reduction in the absolute value of the Seebeck coefficient is related to the lattice distortions induced by the association of multiple nitrogen vacancies in contiguous positions. The transmission function of BT 333\textit{vv} significantly differs from that of BT 333\textit{v} in the region of the lower-energy conduction band. Furthermore, the BT 333\textit{vv} function exhibits a spike within the fundamental gap, where it reaches one quantum of transmittance.

In Figure \ref{fig:TL_cond_333}, we report a comparison between the calculated conductance functions of two (3×)3×3\textit{+s} systems: a TL system, composed of the defected 333\textit{vv} structure as the central conductor with semi-infinite repetitions of pristine 333 cells as the source and drain leads, and a BT system, composed of an infinite repetition of pristine 333 conventional cells. The results highlight the impairment in electronic transport caused by lattice distortions in the defected system. This effect is counterbalanced by the increase in carrier concentration resulting from \textit{n}-type doping from vacancies.

In Figures \ref{fig:cond_Seebeck_155O} and \ref{fig:cond_Seebeck_333O} we present the resulting transmission functions and Seebeck coefficients profiles for the point-defected BT systems with central O impurities in place of N sites. The results of our calculations show that this kind of impurities have negligible effects on the profiles of Seebeck coefficients when compared to the effects of vacancies on N sites. This holds true whether the O impurities are isolated (333O) or packed in contiguous positions (155O).


In Figures \ref{fig:cond_Seebeck_155v} and \ref{fig:cond_Seebeck_155O}, alongside the pristine and BT single‐point‐defected (1x)5x5+s systems, we also present the calculated transmission functions and Seebeck coefficient functions of systems combining—in contiguous positions and BT configuration—{stacking faults} and point defects (155\textit{v}+{\textit{{sf}}} and 155O+{\textit{{sf}}}). We observe that the presence of a {stacking mismatch} produces a pronounced decrease in the Seebeck coefficient profile in both cases, comparable to that induced by multiple vacancies at contiguous N sites. This allows us to understand that the effects related to vacancies association (\textit{v}+\textit{v}) are primarily electronic in nature when compared to the effects of oxygen associations (O+O or O+{\textit{{sf}}}), since they manifest similarly whether or not the vacancy is paired with a stacking {fault}, whereas the impact of an oxygen impurity adjacent to a {planar} defect (O+{\textit{{sf}}}) is predominantly structural in nature when compared to \textit{v}+\textit{v}, as its profile reduction of the Seebeck function disappears in the absence of the {stacking mismatch}.


In Figure \ref{fig:333_Seebeck}, we report the calculated Seebeck functions for selected BT 3×(3×3)\textit{+s} ScN NW systems. These functions were obtained {after a rigid shift of the conduction bands in order} to reproduce the 0.9 eV experimental gap. It can be observed that this procedure yields numerically reasonable results when compared with experimental values. However, this approximation overestimates the distance between the calculated chemical potential and the peak values of the Seebeck functions. Indeed, reducing the cross-section of the NW models leads to increasingly higher absolute peak values in the calculated Seebeck functions, and similar effects are observed when a rigid shift is applied to the energy bands.

For this reason, the Seebeck functions calculated without the application of a rigid shift retain a greater physical significance in their analytical shape. In Figure \ref{fig:333_Seebeck} we can still observe a reduction in the absolute value of the Seebeck coefficients for systems with vacancies at N sites. Although these differences between the single- and double-vacancy defected systems are still present, they are less pronounced here than those observed between the peak values of the Seebeck functions calculated without applying a rigid shift (Figure \ref{fig:cond_Seebeck_333v}).

{To summarize, we have identified two notable categories of defect structures that produce a significant effect on the shape of the Seebeck functions: \textit{v}+\textit{v} and O+\textit{{sf}}. These two categories of defects {might} be responsible for part of the experimental variability in the Seebeck coefficient reported in the experimental literature \cite{10.1063/1.4801886,doi:10.1021/acsaem.2c00485,doi:10.1021/acsaem.2c01672,MORECHEVALIER2025100674,bouteiller2025improving} and will be further {investigated} in the following subsection.}

\ 
\ 
\

{\subsection{Electronic conductivities}}

{The band structure of the 155\textit{v} model (Fig. \ref{fig:bands_155O_155v}) suggests that the presence of \textit{v}+\textit{v} configurations can lead to a substantial increase in electronic conductance. After introducing — using the same procedure described above — a rigid shift of the conduction bands to reproduce the experimental 0.9~eV band gap, we computed electric conductivities $\sigma(\mu)$ as functions of the chemical potentials from the different conductance functions G of our NW models:}

\begin{equation}
  \sigma (\mu) = G (\mu) \cdot \frac{L}{A} \cdot MP
\end{equation}

where $L$ is the length of the central conductor, $A$ is the NW cross-sectional area, and $MP$ is a model parameter obtained by fitting experimental data. {The Landauer approach, with the inclusion of a finite mean free path for backscattering, has been shown to be formally equivalent to the Boltzmann transport equation within the relaxation time approximation \cite{10.1063/1.3291120}}. This heuristic $MP$ parameter accounts for (i) the effect of a finite mean free path in real materials relative to the ballistic transport assumed in our simulations, (ii) the difference in dimensionality between our nanoscale models and the larger experimental thin-film system , (iii) the approximate correction of the band gap we employed, and (iv) the compensation for the partially arbitrary choice of an auxiliary function in the model (which will be described later in the text).

We relate the electronic conductances of our NW models to experimental electric conductivity data available in the literature. In Figure~\ref{fig:155_model}, panel \textit{b}, it can be observed that the BT 155\textit{v} systems exhibit the highest conductivity among the systems we have examined. We associate this model with an hypothetical real ScN system characterized by a strong predominance of N-site vacancies occurring in contiguous positions. The experimental ScN sample from Ref.~\onlinecite{MORECHEVALIER2025100674} exhibits some of the highest electric conductivity values ever reported for ScN thin films.


We assume that a strong presence of structural vacancies at N-sites, associated in contiguous positions, is at the origin of the high experimental conductivity. It should be noted that the association of vacancies at N-sites in contiguous positions is not thermodynamically favored. We treat surface defects with the idea to describe the deposition process, and in order to quantify the energy involved, we performed calculations of the formation enthalpies of isolated surface vacancies at N-sites in a ScN slab structure, as well as the formation of associated defects {\footnote{We calculated the formation energies of associated defects in 6-atomic-layer slab structures by means of DFT as described in section \ref{sub:detail}, using a 3×3 k-point grid and exploiting space-group symmetries where possible. The periodic unit used for these calculations had in-plane dimensions of 3×3 conventional simple cubic cells. Atomic layer 1 was kept fixed during the relaxation calculations, and defects were inserted in layers 4 and 6 at the in-plane center of the periodic unit, in axial positions perpendicular to the slab.}}. The formation enthalpy of two contiguous vacancies results higher by 8.92\% compared to that of two isolated vacancies (5.44~eV per single isolated vacancy). However, the formation of a single vacancy at N-site is itself thermodynamically unfavorable, and the first studies on the stability of nitrogen vacancy defects~\cite{Porte_1985} argued that these are not energetically favored and were therefore expected not to be significantly present in ScN lattices. More recent experimental reports have clearly shown, instead, that nitrogen vacancies are prevalent, especially under scandium-rich growth conditions~\cite{ALBRITHEN2002345,Smith20011809}. Although these conditions are not easily achieved with sputtering techniques, this suggests the existence of complex mechanisms for the stabilization of nitrogen vacancies that are not yet fully understood. For instance, the formation of void regions in the bulk structure of ScN has been shown to facilitate a higher concentration of \textit{n}-type impurities in their vicinity~\cite{PhysRevMaterials.5.084601}, and other theoretical studies have demonstrated a negative formation energy for surface N-site vacancies under Sc-rich conditions~\cite{PhysRevB.65.161204}.


These reasons moved to our assumption and therefore we derived a single model parameter $MP$, which we used for all the systems presented in Figure~\ref{fig:155_model}, by fitting the BT 155\textit{v} conductivity to the experimental data reported in Ref.~\onlinecite{MORECHEVALIER2025100674}, corresponding to the sample referred to in that work as ScN (here we denote it as Sample I). By intersecting a model-internal theoretical auxiliary function, which we report in Figure~\ref{fig:155_model}, panel \textit{b} (155\textit{v} aux.), we derived the chemical potentials of that sample at various temperatures from the experimental Seebeck coefficients reported in the same work~\cite{MORECHEVALIER2025100674}. As auxiliary function, we used the theoretical Seebeck function of each NW model itself, calculated at a temperature of 450 K (this choice is partially arbitrary and is compensated by the derivation of a suitable $MP$, as previously discussed). We obtained the best possible fit of the model by associating the auxiliary function of the 155 model with the experimental sample labeled ScN-T in Ref. \onlinecite{MORECHEVALIER2025100674} (here denoted Sample II), the auxiliary function of the BT 155\textit{v} model with the experimental sample labeled ScN in Ref. \onlinecite{MORECHEVALIER2025100674} (Sample I, used to fit $MP$), and the auxiliary function of BT 155O+{\textit{{sf}}} with the experimental sample examined in Ref. \onlinecite{10.1063/1.4801886} (here denoted Sample III). In Figure \ref{fig:155_model}, we report these three auxiliary functions and the calculated conductivities comparing with experimental values from Refs. \onlinecite{10.1063/1.4801886} and \onlinecite{MORECHEVALIER2025100674}.

We chose to compare the two models of defected structures noted before that present the most interesting features when comparing their Seebeck function profiles with that of the pristine 155 model, namely: BT 155\textit{v} and BT 155O+{\textit{{sf}}}. The compared experimental samples (Refs. \onlinecite{10.1063/1.4801886} and \onlinecite{MORECHEVALIER2025100674}), deriving from different experimental procedures, differ widely from each other in their characteristics, presenting a wide variability in Seebeck coefficients, electric conductivities, and their variance with temperature. We notice a good agreement between the experimental conductivity values measured for the ScN-T sample in Ref. \onlinecite{MORECHEVALIER2025100674} (Sample II) and the theoretical functions calculated using the NW model 155O+{\textit{{sf}}} (we mention a slightly lower, yet still good, agreement when using the auxiliary function of the 155O model, not reported in figure).

We can therefore describe, within a coherent framework, an interaction between two independent effects that determine the different thermoelectric behaviors of the ScN and ScN-T samples in Ref. \onlinecite{MORECHEVALIER2025100674} (Sample I and II). The different experimental procedures lead, on one hand, to a reduction in \textit{v}+\textit{v} defects in ScN-T compared to ScN. This reduction underlies the strong increase in the absolute value of the Seebeck coefficient observed in Ref. \onlinecite{MORECHEVALIER2025100674} and also contributes to the decrease in electrical conductivity. On the other hand, an additional detrimental effect on electronic transport arises from a greater presence of O+{\textit{{sf}}} defects.

Further complexity is detected when analyzing the experimental sample from Ref. \onlinecite{10.1063/1.4801886} (Sample III). This sample also exhibits high conductivity values, so it is reasonable to expect its experimental values to match the highest conductivity function, namely the one calculated using the 155\textit{v} model. However, in order to achieve good agreement, a different auxiliary function must be used in this case. The best match is obtained when using an auxiliary function calculated from the BT 155O+{\textit{{sf}}} system. The choice of a different auxiliary function accounts for structural effects that may influence the Seebeck coefficient but have little impact on the conductivity. These considerations suggest that different defected structures can coexist within the same sample, acting independently on the electric conductivity and the Seebeck coefficient, with possible contrasting effects.

At this point, it should be noted that while the 155\textit{v} model has a physically unique structure determined by a structural optimization calculation, the structure used for the 155O+{\textit{{sf}}} model involves a greater degree of arbitrariness in the way the planar mismatch {was introduced as a stacking fault}. As previously discussed, the thermoelectric properties of the 155O+{\textit{{sf}}} model are highly structure-dependent. From an analytical perspective, the shape of its auxiliary function is well suited to fit the model. However, to achieve better numerical agreement, one would need to use the Seebeck function calculated at 600 K as auxiliary function (instead of the one at 450 K, which we use). We instead choose to preserve the internal consistency of the model by calculating all auxiliary functions in the same way. As a result, this set of points yields a lower agreement between theory and experiment compared to the others. However, the reasons for this outcome are clear, as the 155O+{\textit{{sf}}} defect category inherently includes a greater amount of structural variability.

{To summarize, based on the results of our theoretical calculations, we derived a heuristic model that relates our simulated atomistic systems to the experimental conductivities reported in Ref. \onlinecite{MORECHEVALIER2025100674} and \onlinecite{10.1063/1.4801886}. We used the data of Sample I (ScN, Ref. \onlinecite{MORECHEVALIER2025100674}) to fit the model. We obtained good agreement with the experimental conductivities of Sample II (ScN-T, Ref. \onlinecite{MORECHEVALIER2025100674}). A slightly weaker agreement is found with Sample III (ScN, Ref. \onlinecite{10.1063/1.4801886}).}


\  
\   
\

\section{Conclusions}

{We wondered what the microscopic origins of the experimental variability in thermoelectric response of ScN thin films might be. Knowing that the thermoelectric properties of ScN are strongly influenced by stoichiometric variability, structural imperfections, and chemical impurities, we decided to model some of these structural features to understand how they act in modifying the electronic transport, and consequently the thermoelectric response.}

{We employed a theoretical framework suitable for studying in detail these atomistic effects on quantum transport. We approached the problem by classifying the possible structural defects and impurities based on their symmetry and chemical nature, and we systematically analyzed those most likely to occur under the experimental conditions typically used.}

From the point of view of defect symmetry, we considered isolated point defects, multiple contiguous point defects, and point defects associated with planar mismatches {(stacking faults)}, while from a chemical perspective we studied the presence of oxygen or structural vacancies replacing nitrogen atoms in the lattice. We used the Landauer approach to determine the effect of each of these different microscopic structures on the emergent thermoelectric properties of the whole material.

{Thermoelectric conversion is a complex phenomenon and electronic transport is certainly influenced, through electron-phonon scattering, by extended defects such as nanovoids, grain
orientation, substrate-induced strain. Therefore our theoretical approach cannot account for
the full extent of the experimental variability. Nevertheless, electronic transport is also fundamentally related to the individual quantum states and to the microscopic regions where the single electrons flow.}

We identified two notable categories of structures whose presence has a considerable impact on the Seebeck coefficient and electrical conductivity at the {microscopic} level and which, if significantly present in the material, could {partially} account for the experimentally observed behavior.

Specifically, we refer first to the \textit{v}-\textit{v} structures, where two or more structural vacancies replace nitrogen atoms at adjacent conventional cells (i.e., N-site vacancies separated by only one Sc atom), which can significantly reduce the absolute value of the Seebeck coefficient while enhancing the electrical conductivity. Second, we refer to the O+{\textit{{sf}}} structures, where an oxygen atom replaces a nitrogen lattice atom, and a {stacking fault} is immediately adjacent, involving a displacement of the atomic layer following the one where the substitution occurs. This second category of structures has potentially opposite effects with respect to the firts one, detrimenting the conductivity while enhanching the Seebeck coefficient.

{Since \textit{n}-type defects tend to accumulate in the vicinity of nanovoid regions in the ScN lattice, the proximity of these structures could allow \textit{v}-\textit{v} and O+{\textit{{sf}}} formations to reach a concentration sufficient to produce an appreciable effect on the thermoelectric response of experimental samples, offering a rationale for part of the observed experimental variability.}

With this work, we aim to contribute to the theoretical understanding of the microscopic mechanisms governing electronic transport in ScN. Given the complexity of this matter, the problem of optimizing the thermoelectric performance of ScN nanostructures by independently tuning their thermal and electrical transport remains open. Identifying a rational strategy—by applying theoretical insights to the design of experimental work—is a fundamental step for future progress.

\ 
\

\section*{Acknowledgements}

This work was supported by the Ministry of Education, Youth and Sports of the Czech Republic through the e-INFRA CZ (ID:90254) and project QM4ST, No. CZ.02.01.01/00/22\_008/0004572 and also by Czech Science Foundation grant No. 23-07228S.

\
\

\section*{Declaration of Competing Interest}

The authors declare that they have no known competing financial interests or personal relationships that could have appeared to influence the work reported in this paper.

\
\

\section*{Authorship contribution statement}

L. Cigarini: Conceptualization, Methodology, Investigation, Formal analysis, Visualization, Writing - Original Draft. U.D. Wdowik: Conceptualization, Resources, Validation, Writing - Review \& Editing. D. Legut: Conceptualization, Resources, Validation, Writing - Review \& Editing, Supervision, Project administration, Funding acquisition.

\bibliography{biblio}

\end{document}